\documentclass[prd,eqsecnum,apsi]{revtex4}

\usepackage{graphicx}
\usepackage{latexsym}
\usepackage{amsmath}
\usepackage{amssymb}

\newcommand{\be}{\begin{equation}}
\newcommand{\ee}{\end{equation}}
\newcommand{\bea}{\begin{eqnarray}}
\newcommand{\eea}{\end{eqnarray}}
\newcommand{\bfig}{\begin{figure}}
\newcommand{\efig}{\end{figure}}

\begin{document}

\title{\Large \bf A critical analysis of thermodynamic properties of  braneworld black holes in anti-de Sitter spacetime}

\author{Ratna Koley $^{1, 2}$\footnote{E-mail:ratna.koley@rediffmail.com},
Joydip Mitra $^1$ \footnote{E-mail:tpjm@iacs.res.in}, Supratik Pal
$^3$ \footnote{E-mail:supratik\_v@isical.ac.in} and Soumitra
SenGupta $^1$ \footnote{E-mail:tpssg@iacs.res.in}}
\affiliation{$^1$Department of Theoretical Physics and Centre for
Theoretical Sciences, \\
Indian Association for the Cultivation of Science, \\
Kolkata  700 032, India \\
$^2$Department of Physics,  Bethune College, \\ 181 Bidhan
Sarani, Kolkata 700 006, India \\
$^3$Physics and Applied Mathematics Unit, \\
Indian Statistical Institute, \\
203 B. T. Road, Kolkata 700 108, India}

\begin{abstract}
Thermodynamic properties of
Schwarzschild-Anti de-Sitter (Sch-AdS) and
Reissner-Nordstr\"{o}m-Anti de-Sitter (RN-AdS) blackholes in 3+1 dimensional spacetime 
are studied critically with special reference to the warped braneworld black holes with non-vanishing cosmological constant
on the brane.
Explicit dependence of
the thermodynamic variables on the parameters of the braneworld
model such as the induced three brane cosmological constant as well as the bulk cosmological constant have been determined.
Hawking-Page phase transition has been discussed for both
Sch-AdS and RN-AdS black holes. At the phase transition point it is shown that the parameters 
mass, charge and cosmological constant get related by
an inequality relation which originates from the background warped geometry model. 
\end{abstract}

\maketitle

\section{Introduction}

Standard model of elementary particles, despite it's success in
explaining physics upto a scale near TeV, has the well known fine
tuning/gauge hierarchy problem in connection with the only scalar in
the theory namely Higgs. Supersymmetry resolves this problem at the
expense of introducing large number of superpartners in the theory,
none of which has been detected so far. Alternatively theories of
extra dimensions were developed recently. The warped geometry model
proposed by Randall and Sundrum was particularly successful in
resolving the gauge hierarchy problem without bringing in any
intermediate mass scale in the theory. In this model the only extra
dimension in a 5 dimensional anti-de Sitter bulk space-time is
compactified on a $S_1/Z_2$ orbifold and two 3-branes are placed at
the two orbifold fixed points at $y = 0$ and $y = \pi$ separated by a distance $r$. Because of the bulk gravity it turns
out that the two branes have hierarchial mass scales even when the
bulk cosmological constant and brane separation modulus are at the
Planck scale. In the effective 4D theory the lower energy ( TeV )
brane is flat with zero cosmological constant. Generalizing this
model further, it has been shown recently that the gauge hierarchy
problem  can be resolved in the case of a non-flat 3-brane
\cite{genrs} also, contrary to the Randall-Sundrum model with flat
3-brane \cite{rs}. The metric for the above spacetime is given by :
\be ds^2 = e^{- 2 A(y)} g_{\mu\nu} dx^{\mu} dx^{\nu} +r^2 dy^2
\label{metric} \ee
where %
\be \label{wfads} e^{-2A} = \omega^2 \cosh^2 \left(\ln \frac
{\omega} c_1 + ky \right) \nonumber \ee

Here $\Omega$ is the brane cosmological constant of the (3 + 1)-
spacetime and $\omega^2=-\frac{\Omega}{3 k^2} $ and  Depending on the signature of $\omega^2$ the brane can be
either asymptotically AdS or dS . $c_1=1+\sqrt{1-\omega^2}$ and $k$ is related to the Bulk 
cosmological constant($\Lambda$) as $k \sim \sqrt{-\Lambda/12 M^3}$. The matter content on the four
dimensional hypersurface can give rise to Schwarzschild-anti de
Sitter (Sch-AdS henceforth) or  Reissner-Nordstr\"{o}m-anti de
Sitter (RN-AdS) blackholes on the brane for negative values of
$\Omega$. Note that in all the cases the hierarchy problem is
solved in the gravitational weak field limit. There are several
possibilities that can occur in the above braneworld scenario. The
brane tension can be either positive or negative for  $\omega^2 < 0
$. For negative brane cosmological constant there is an upperbound
on the value of $\omega^2$ for which the hierarchy issues can be
resolved. Moreover it has been shown  that the $\omega^2$ plays a
significant role in determining  the stability of the spacetime
\cite{genrsstab} and in localizing a  bulk fermion near the brane
\cite{genrsloc}.

In the present article our primary intention is to investigate the
implications of the presence of a brane cosmological constant
$\omega^2$ to the thermodynamics of four dimensional asymptotically
AdS black holes. We show that the constraints that arise from the
requirement of the resolution of the gauge hierarchy problem have
interesting implications in the black hole physics on the brane. The
quest for a consistent description of black hole physics and
gravitational collapse in braneworlds \cite{rs2} has been a
challenge to theoretical physicists. 
The first attempt in this direction was to provide 
a black string solution \cite{blstring}, with extended singularity in higher dimensions, which turned out to be
neither stable nor localized on the brane. Leter on, it was found that the simplest
localized black hole solution  on the brane is
Reissner-N\"{o}rdstrom type with the 
usual charge replaced by a geometric quantity called `tidal charge' \cite{tidal}.  Subsequently, black hole solutions with  non-vacuum brane
and the black hole intersecting the bulk \cite{seahra, adsbh} and for charged rotating black holes were
obtained \cite{chargerot}. Some good results in this direction can also  be found in \cite{revbh, bhrecent}. 
Oppenheimer-Snyder type \cite{os} gravitational
collapse of spherically symmetric objects was also studied in
\cite{coll1} and the new results were formulated as a no-go theorem leading to a non-static exterior
for the collapsing sphere on the brane. Subsequent generalizations 
that include, among others,  the Gauss-Bonnet gravity for the bulk spacetime have been carried out in 
\cite{coll2, collgen}. However, later on, it was demonstrated in \cite{sucol} that  a
static exterior of the Schwarzschild can be obtained by relaxing
certain assumptions, thereby re-establishing Birkhoff's theorem for
braneworld black holes as well. All these results lead to the 
conclusion that we are yet to make a strong conclusive comment about the physics of black holes in the braneworld scenario.

On the other hand, in the general relativistic framework, different issues related to
thermodynamics of black holes  have been addressed over the years.
The reader may go through \cite{tdrev} for a useful review of the
subject. Though the thermodynamics of asymptotically anti de Sitter black holes
have always drawn much attention \cite{hawking}, there has been a renewed interest in this area mainly due to their string
theoretical connections \cite{tdstring}. Some recent works in this
direction include investigation for quasinormal modes for such black
holes. For example, \cite{tdscads} deal with  Sch-AdS black holes
whereas \cite{tdrnads} discuss the case for RN-AdS black holes.
However, black hole thermodynamics is somewhat less explored  in the
braneworld scenario. Although there have been some works in the
literature that deal with certain thermodynamic features of
braneworld black holes \cite{tdbrbh}, they are mainly developed in
the context of  single brane scenario. Our aim in this article is to
extend the analysis in the framework of curved two-brane scenario
proposed in \cite{genrs}. As it will turn up, our analysis itself
has certain interesting features that worth studying. The salient
features of our analysis are the following: The study of black hole
thermodynamics for  curved two-brane models is the first of its kind
in two-brane context. To this end, we have obtained exact analytical
solutions for different thermodynamic quantities, both for Sch-AdS
as well as for RN-AdS black holes. We have further established
certain link of the major black hole parameters (mass, charge and
cosmological constant to be precise) for asymptotically AdS black
holes on the brane to the bulk cosmological constant via
thermodynamic analysis.
This interpretation in turn leads to the interesting finding that an
upper bound for the black hole mass results from an upper bound for
the cosmological constant in this particular warped braneworld
context. In a nutshell, here we have a curved brane scenario which
not only addresses issues like gauge hierarchy, stability and
fermion hierarchy, but also  provides important insights to the
thermodynamics properties of asymptotically AdS black holes on the
brane.


\section{Schwarzschild-AdS black hole}
The four dimensional brane metric representing a Sch-AdS spacetime
is given by \be ds_{(4)}^2=-\left(1-\frac{2 m/M_{Pl}^2}{r}+k^2
\omega^2 r^2 \right) dt^2+ {\frac{dr^2}{\left(1-\frac{2
m/M_{Pl}^2}{r}+k^2 \omega^2 r^2 \right)}} + r^2 d\Omega^2
\label{bhmetric} \ee

with $\omega^2=-\frac{\Omega}{3 k^2} $ where $\Omega$ is the induced brane-cosmological constant and $m$ corresponds to the
mass parameter of the Sch-AdS blackhole. For $\Omega > 0$ we obtain
Sch-AdS solution on the brane. Parameter $k$ is related to the bulk cosmological constant($\Lambda$) and hence encodes
the bulk effects on the brane black hole parameters.
Considering the above  metric in Eq.
(\ref{bhmetric}) one can easily find the expression for the horizon.
There exist three roots for the equation $g_{00}(r) = 0$, out of which
two  are complex and only one is real. The real root determines the
horizon $r_H$ which is explicitly given by

\be \label{rh} r_H=-\frac{1}{3^{\frac{1}{3}} \left(\frac{9 m k^4
\omega ^4}{M_{Pl}^2} +\sqrt{3} \sqrt{k^6 \omega ^6+ \frac{27 m^2 k^8
\omega ^8}{M_{Pl}^4} }\right)^{\frac{1}{3}}}+\frac{\left(\frac{9 m
k^4 \omega ^4}{M_{Pl}^2} +\sqrt{3} \sqrt{k^6 \omega ^6+ \frac{27 m^2
k^8 \omega ^8}{M_{Pl}^4} }\right)^{\frac{1}{3}}}{3^{\frac{2}3{}} k^2
\omega ^2} \ee

The requirement, $r_H > 0$ gives rise to the condition $(m/M_{Pl}^2)
k^7 \omega^7 > 0$ which is automatically satisfied and puts no extra
constraints on the parameters of the theory. This implies that there
will always exist a horizon for the Schwarzschild-AdS metric.
In Fig. (\ref{rws}) we plot the variation of $r_{H}$ with
respect to $\omega$.

\bfig[h]
\includegraphics[height = 5 cm, width = 7 cm]{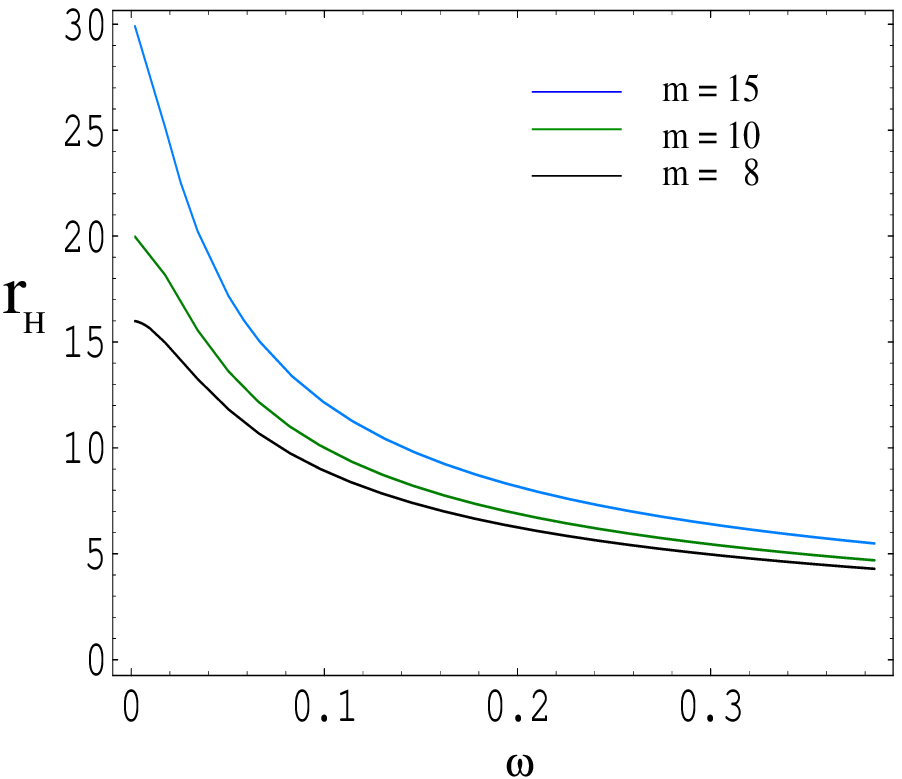}
\caption[]{$ r_{H}$ is plotted against $\omega$ for different values
of m} \label{rws} \efig

It is interesting to note that if we consider $\omega^2$ to be very
very small then the expression for horizon in the leading order is
given as,
\be r_H=2 (m/M_{Pl}^2)-8 (m^3/M_{Pl}^6) k^2 \omega^2 \ee
For $\Omega=0$,  $r_H=2 (m/M_{Pl}^2)$ which is same as an
asymptotically flat Schwarzschild black hole in four dimensions.
The non-zero cosmological constant therefore reduces the horizon radius from that
of the Schwarzschild value. 
We now estimate other thermodynamic quantities such as the Hawking
temperature ($T_H$), Entropy($S$) and the specific heat $C_v$ for
this class of black holes. For brevity let us introduce a function
$P(\omega)$ as, 
\be 
P(\omega) = \left(9 (\frac{m}{M_{Pl}^2}) k^4
\omega ^4+\sqrt{3} \sqrt{k^6 \omega ^6+27 (\frac{m^2}{M_{Pl}^4}) k^8
\omega ^8}\right)^{1/3} 
\ee 
In what follows we shall express the quantities in terms of this newly-introduced function, which contains the brane
cosmological constant $\omega$ as well as the bulk cosmological constant through the parameter $k$.

The Hawking temperature for the above metric (\ref{bhmetric}) turns
out to be

\be \label{T1} T=\frac{-3^{\frac{1}{3}} k^2 \omega
^2+\left(P(\omega)\right)^{\frac{2}{3}}+\frac{9 (m/M_{Pl}^2) k^4
\omega ^4 \left(P(\omega)\right)}{\left(-3^{\frac{1}{3}} k^2 \omega
^2+\left(P(\omega)\right)^{\frac{2}{3}}\right)^2}}{2 ~
3^{\frac{2}{3}} \pi  \left(P(\omega)\right)^{\frac{1}{3}}} \ee

The temperature profile in Eq.(\ref{T1}) has been plotted in 
figure {\ref{T1}}. 
\bfig[h]
\includegraphics[height = 5cm, width = 7 cm]{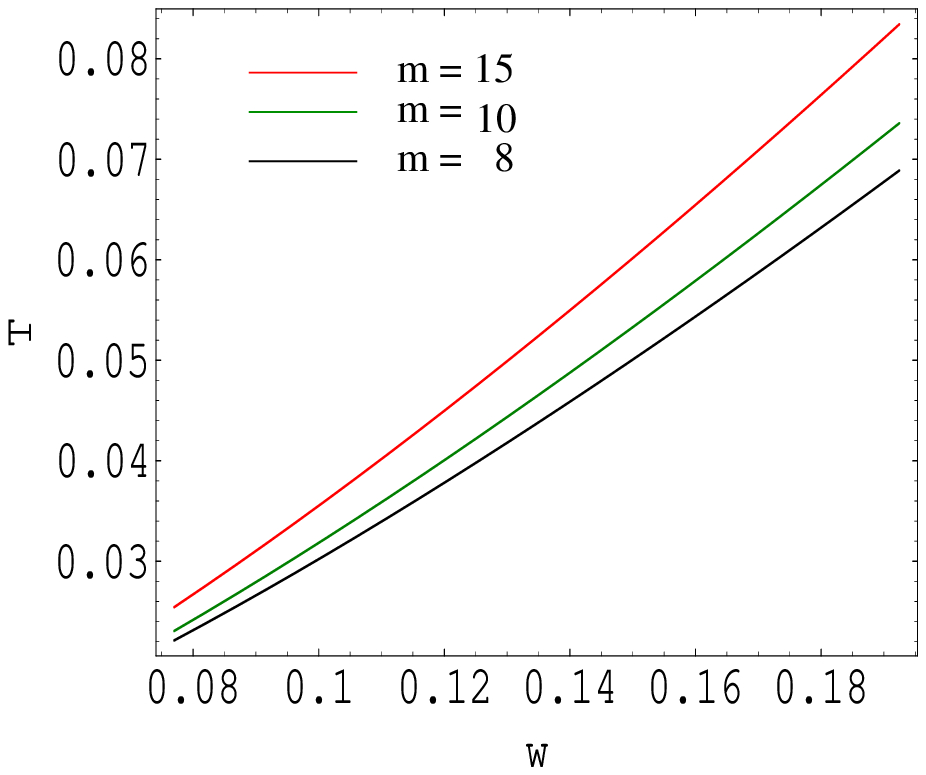}
\caption[hawking]{$T$ is plotted against $ \omega $ for different
values of mass} \label{T1} \efig The Hawking temperature turns out
to be a increasing function with respect to $ \omega$. As it has been shown earlier that
the horizon radius decreases with increasing cosmological constant, the Hawking temperature, which
is inversely related to the Hawking radius for purely Scwarzschild case, therefore predictably
grows with cosmological constant.

Employing the thermodynamic relations we obtain the expression for the
Bekenstein-Hawking entropy as

\be S=\pi  \left(-\frac{1}{3^{1/3}
P(\omega)^{1/3}}+\frac{P(\omega)^{1/3}}{3^{2/3} k^2 \omega
^2}\right)^2 \label{s1} \ee

The variation of Entropy with respect to $\omega$ (in units of
$M_{Pl}=1$ and  $k=M_{Pl}$) has been shown in the figure {\ref{S1}}.
\bfig[h]
\includegraphics[height = 5cm, width = 7cm]{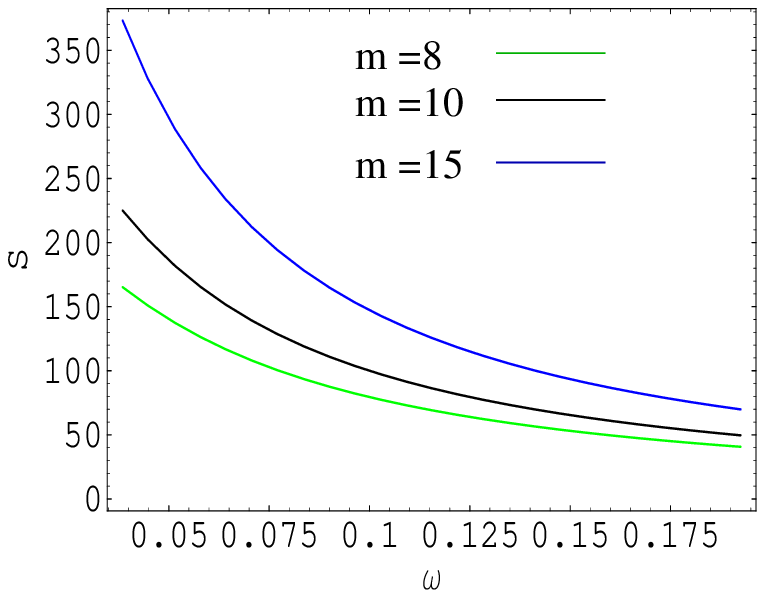}
\caption[]{$ S$ is plotted against $ \omega$ for different values of
m} \label{S1} \efig

The entropy thus falls with increase in the value of the brane
cosmological constant as shown in the figure. For a given value of $
\omega$, the entropy is larger for a larger value of the mass
parameter $m$ as is expected. For large value of $ \omega$, the role
of the mass parameter becomes less significant and curves for
different $m$ slowly converges.

Finally the expression for the specific heat $C_v$ is found to be,
\bea \label{Cv} C_v & = & \frac{1}{9 k^6 \omega ^6 \left(-4+27
(m^2/M_{Pl}^4) k^2\omega ^2\right)} \left[2 \pi  \left(-243~ 3^{1/3}
(m^3/M_{Pl}^6) k^6 \omega ^6 \left(P(\omega)\right)^{1/3}
\right. \right.  \nonumber \\
& - & 3^{1/6} \sqrt{k^6 \omega ^6+27 (m^2/M_{Pl}^4) k^8 \omega ^8} \left(P(\omega) \right)^{1/3} \left(2 ~3^{2/3}+9 (m/M_{Pl}^2) \left(P(\omega)\right)^{1/3}\right) \nonumber \\
& + & k^2 \omega ^2 \left(27~ 3^{5/6} (m^2/M_{Pl}^4) \sqrt{k^6
\omega ^6+ 27 (m^2/M_{Pl}^4) k^8 \omega ^8}
\left(P(\omega)\right)^{1/3}-2~3^{2/3}
\left(P(\omega)\right)^{2/3}\right) \nonumber \\
& + & \left.\left.3 k^4 \omega ^4 \left(4 + 9 ~3^{1/3} (m/M_{Pl}^2)
\left(P(\omega)\right)^{1/3} \left(1+2~ 3^{1/3} (m/M_{Pl}^2)
\left(P(\omega)\right)^{1/3}\right)\right)\right)\right] \eea The
plot has been shown in Figure \ref{cv1}.

\bfig[h]
\includegraphics[height = 5cm, width = 7cm]{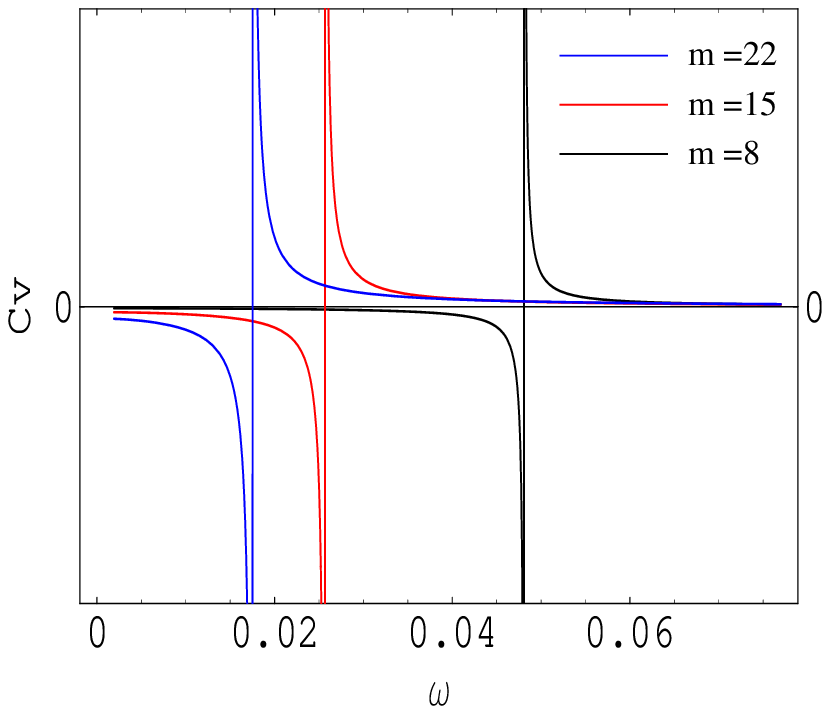}
 \caption[]{$ C_v$ is plotted against $ \omega$ for
different values of m} \label{cv1}
\efig


In contrary to the ordinary Schwarzschild metric where $C_v$ is
always negative and no phase transition is possible at any
temperature, here we obtain the possibility of phase transition. The
phase transition ( for a given value of m) occurs at a particular value of
$ \omega$. The corresponding value of the temperature can be
determined from the $ \omega$ versus $T$ plot given in the previous
figure. For larger value of $m$ the phase transition takes place at a lower value of
the cosmological constant and hence at a lower value of temperature. 

Therefore while the Sch-black hole
in a flat space-time intrinsically unstable, an Ads-Sch black hole
can become stable at a temperature when it's specific heat becomes
positive. From the expression of specific heat in Eq. (\ref{Cv}) one
can see that at a critical value $m/M_{Pl}^2=\frac{2}{3 \sqrt{3} k
\omega}$, the specific heat blows up. The entropy at that point is
found to be positive. It is apparent from the Fig. (\ref{cv1}) that
above the critical value of the mass the specific heat is found to
be positive. Therefore in AdS-Schwarzschild case depending on the
values of the mass and the cosmological constant we can have a stable
black hole.  


\section{Reissner-Nordstrom -AdS black hole}

We now focus our attention to the thermodynamics properties of the
Reissner-Nordstrom-AdS blackhole. Let us start with the following
metric which when embedded in five dimensions gives a constant
curvature hypersurface.
\bea
& & ds^2  =  -f(r)dt^2+f(r)^{-1}dr^2+r^2 d\Omega^2 \nonumber \\
& &f(r)  =  1- \frac{2 m}{M^2_{Pl}} \frac{1}{r} +
\frac{q^2}{M^2_{Pl}} \frac{1}{r^2} + k^2 \omega^2 r^2 \label{rna}
\eea

The mass and charge of the blackhole is given by m and q
respectively whereas the AdS space curvature is determined by
$\omega$.
The horizons are determined by solving equation $f(r_H) = 0$.In this
case two of the four roots are necessarily complex and we are left
with only two physical (real) roots $r_+$ and $r_-$ which correspond
to the outer and inner horizon. It is customary to consider  the
outer radius $r_+$ as the horizon $r_H$. Using Eq. (\ref{rna}) we
obtain the following expression for the outer horizon.
\be r_H=\frac{1}{2} \surd \text{X}+\frac{1}{2} \surd
\left(-\text{X}- \frac{6}{3~k^2 {\omega ^2}} +\frac{4 m}{M_{Pl}^2
k^2 \omega ^2 \surd \text{X}}\right) \label{rnhor} \ee where we have
used the following definitions:

\bea
X &=& -\frac{2}{3~{k^2 \omega^2}}+A+B \nonumber \\
A &=&\frac{2^{1/3} \left(1+12 (\frac{q}{M_{Pl}})^2 k^2 \omega^2\right)}{3 k^2 \omega ^2 \left(2+108 \frac{m^2}{M_{Pl}^2} k^2 \omega ^2-72 q^2 k^2 \omega ^2+\sqrt{\left(2+108 \frac{m}{M_{Pl}}^2 k^2 \omega^2-72 q^2 k^2 \omega ^2\right)^2-4 \left(1+12 q^2 k^2 \omega^2\right)^3}\right)^{1/3}} \nonumber \\
B &=&\frac{\left[2+108 \frac{m^2}{M_{Pl}^4} k^2 \omega^2-72
(\frac{q}{M_{pl}})^2 k^2 \omega^2+\sqrt{\left(2+108
\frac{m^2}{M_{Pl}^4} k^2 \omega^2-72 (\frac{q}{M_{Pl}})^2 k^2
\omega^2\right)^2-4 \left(1+12 (\frac{q}{M_{Pl}})^2 k^2
\omega^2\right)^3}\right]^{1/3}}{3 2^{1/3} k^2 \omega^2}
\label{rnredif} \eea

From the above expressions in Eq. (\ref{rnhor}) and (\ref{rnredif})
it is apparent that for the existence of a horizon we must have \be
f = -X-\frac{2}{k^2 \omega^2}+\frac{4 m}{{M_{Pl}^2 k^2 \omega^2}
\sqrt{X}} > 0 \label{extremality} \ee

This in turn tells us that the charge and $\omega^2$ gives a bound
to the blackhole mass for the existence of the horizon. This is the
generalization  to the condition that one obtains between mass and
charge for an asymptotically flat Reissner-Nordstrom solution which in turn 
determines the extremality condition when the two horizons coincide. In
fig.(\ref{rH-cm}) we have plotted the variation of the horizon with
respect to  $\omega$ for a fixed charge.

\bfig[h]
\includegraphics[height = 5cm, width = 7cm]{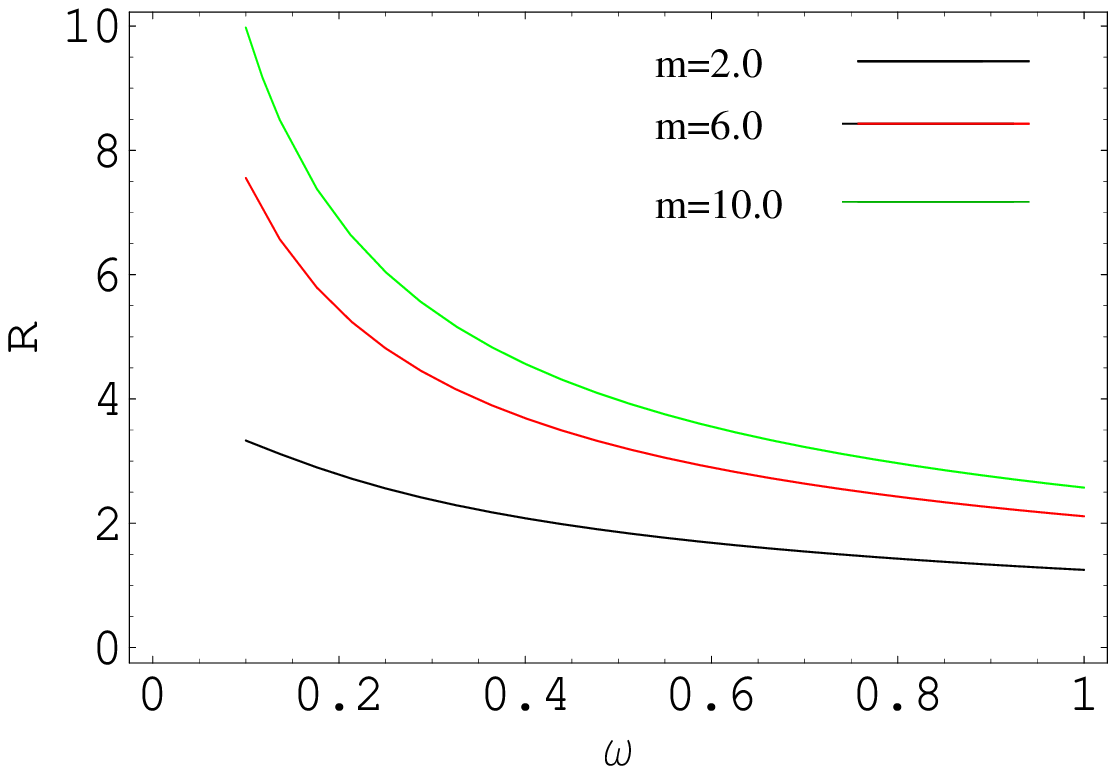}
\caption[Horizon]{$r_H$ is plotted against  $\omega$ and $q =1$}
\label{rH-cm} \efig

Let us determine various thermodynamic quantities for these class of
blackholes and their dependence on the cosmological constant
$\omega$ We first write down the expression of the Hawking
temperature explicitly in terms of mass, charge and  $\omega$ as
follows.

\be T_H=\frac{4 \left(-(\frac{q}{M_{Pl}})^2+\frac{1}{2}
\frac{m}{M_{Pl}^2} \left(\sqrt{\text{X}}+\sqrt{-\frac{2-\frac{4
m}{M_{Pl}^2 \sqrt{\text{X}}}+\text{X} k^2 \omega ^2}{k^2 \omega
^2}}\right)+\frac{1}{16} k^2 \omega ^2
\left(\sqrt{\text{X}}+\sqrt{-\frac{2-\frac{4
m}{M_{Pl}^2\sqrt{\text{X}}}+\text{X} k^2 \omega ^2}{k^2 \omega
^2}}\right)^4\right)}{\pi
\left(\sqrt{\text{X}}+\sqrt{-\frac{2-\frac{4 m}{M_{Pl}^2
\sqrt{\text{X}}}+\text{X} k^2 \omega ^2}{k^2 \omega ^2}}\right)^3}
\ee
    
In the figures below we have shown the variation of blackhole
temperature with respect $\omega$ for a definite charge. (Once again
we have taken $k \simeq M_{Pl}$ and $M_{Pl}=1$).

\bfig[h]
\includegraphics[height = 5cm, width = 7cm]{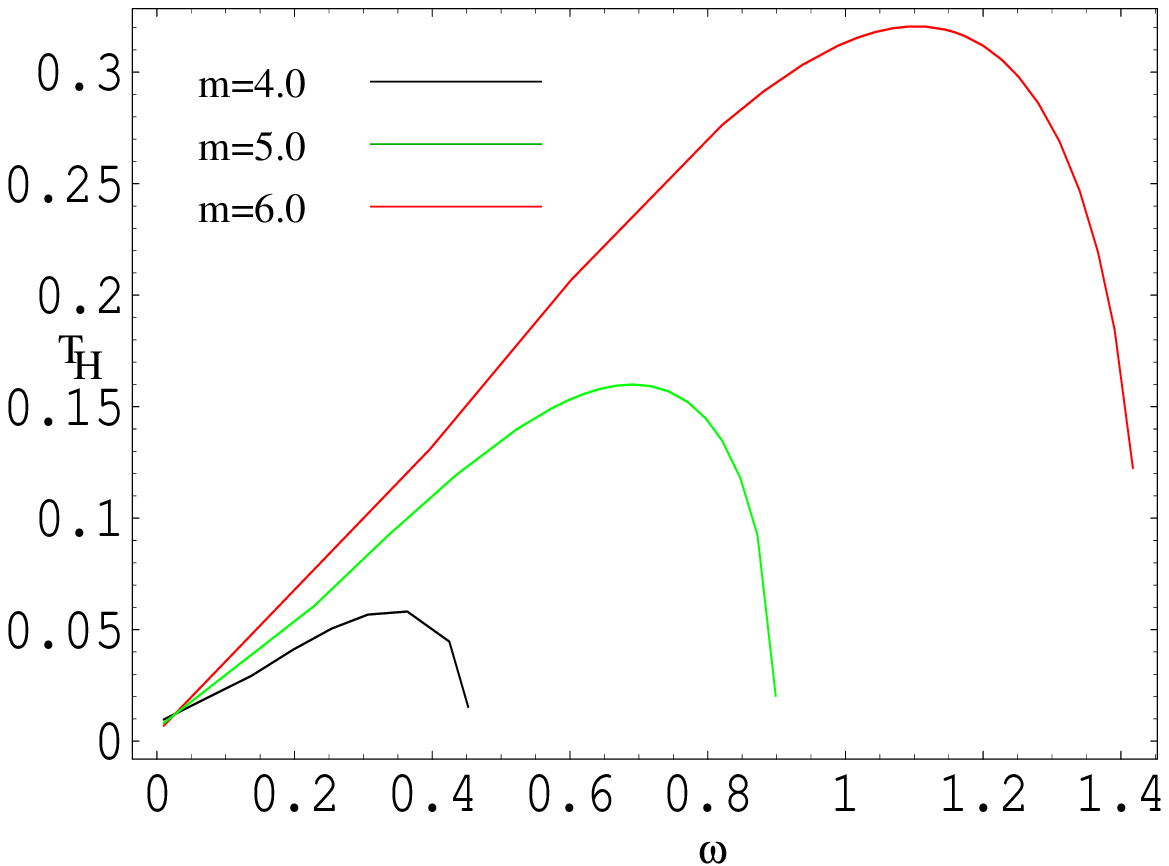}
\caption[Horizon1]{$T_H$ is plotted against $\omega$ for $q=3$}
\label{T1-m} \efig

We have shown earlier that for anti-de Sitter Schwarzschild black hole, the Hawking temperature is
an increasing function of the cosmological constant. It may however be recalled that
in flat space-time  the Hawking temperature corresponding to Reissner -Nordstrom solution is a decreasing function
of the charge. This explains the presence of a turning point in the $T_H - \omega$ ( for a fixed 
value of the charge and mass ) graph in the present scenario where the Hawking temperature
starts decreasing with cosmological constant after certain value which depends on the charge. 
Expectedly the turning point shifts to the right side as the chrage decreases.

The Bekenstein-Hawking  entropy for this black hole is given by the 
expression :

\be S=\frac{1}{4} \pi  \left(\sqrt{\text{X}}+\sqrt{-\frac{2-\frac{4
m}{M_{Pl}^2 \sqrt{\text{X}}}+\text{X} k^2 \omega ^2}{k^2 \omega
^2}}\right)^2 \ee

for a fixed value of $q$  the variation of entropy w.r.t $\omega$ is
plotted in the figure (\ref{S-w}).

\bfig[h]
\includegraphics[height = 5cm, width = 7cm]{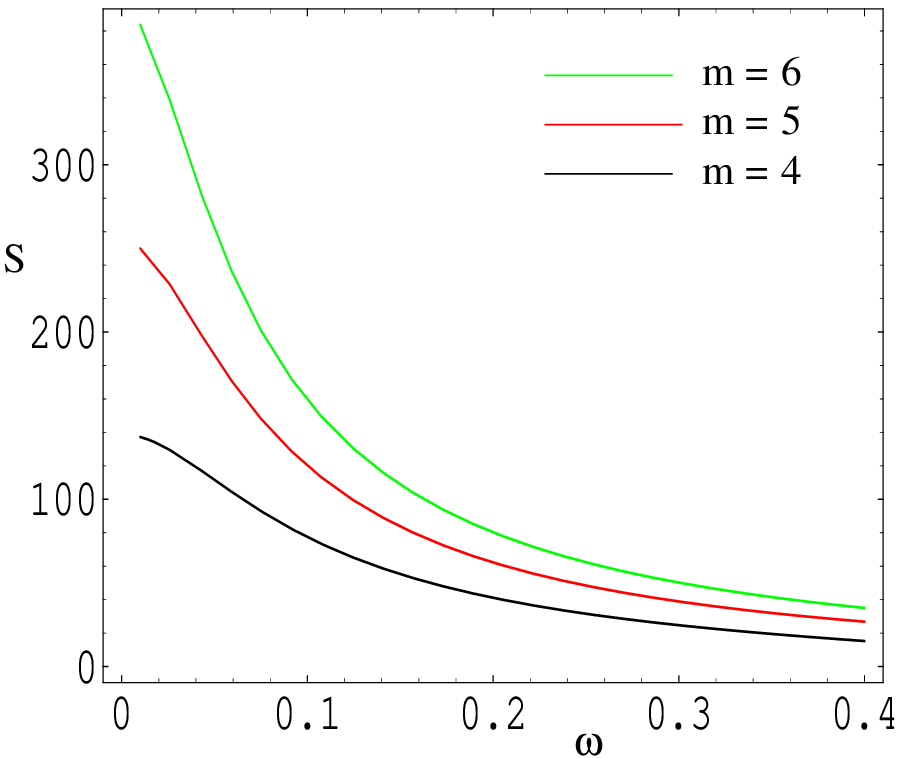}
\caption[Horizon1]{$S$ is plotted against   $\omega$  for  $q=3$ }
\label{S-w} \efig

Finally we obtain  the specific heat of RN-AdS blackhole as

\be C_v=\frac{2 \pi r_H^2(1-\frac{q^2}{M_{Pl}^2 r_H^2}+3 k^2
\omega^2 r_H^2)}{-1+3\frac{q^2}{M_{Pl}^2 r_H^2}+3 k^2 \omega^2
r_H^2} \label{cv-rna} \ee

For a fixed value of $\omega$ and $q$ the variation of specific heat
with respect to mass is plotted below in Fig (\ref{Cv-w}).

\bfig[h]
\includegraphics[height = 5cm, width = 8cm]{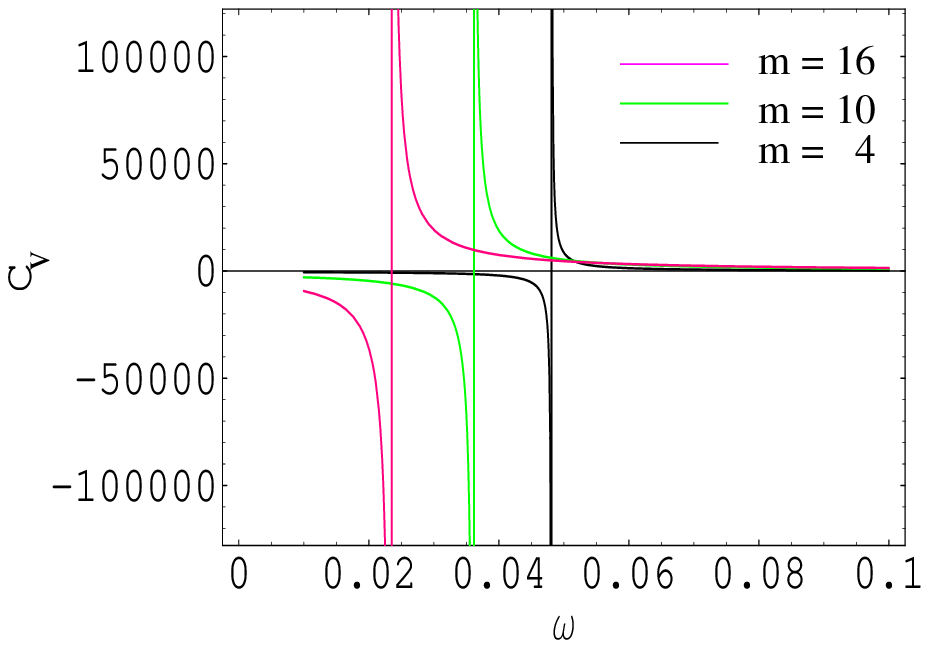}
\caption[Horizon1]{$C_v$ is plotted against $\omega$ for  $q=3$}
\label{Cv-w} \efig

From the expression of the specific heat in Eq. (\ref{cv-rna}) we
see that for $C_v \rightarrow \infty$ the following condition
emerges, \be \frac{q^2}{M_{Pl}^2 y^2}+k^2 \omega^2 y^2=\frac{1}{3}
\ee
 where $y=r_H$. This further leads to the following condition
\be 0<k \omega (q/M_{Pl}) <\frac{1}{6} \ee

So we see that the reality condition for the horizon also determines
the discontinuity of the specific heat. From the graph it is clear
that we have Hawking-Page phase transition in RN-AdS black hole. Since at the
blowing up point the entropy is positive, the phase transition is
necessarily of second order.

\section{Summary and outlook}
This work brings out in details the exact correlations amomg the various parameters like mass, charge,
cosmological constant in the context of thermodynamic properties of black holes in anti-de Sitter
space time. Withour resorting to any approximation our analysis reveals the interesting behaviours
of thermodynamics variables like Hawking temperature, entropy, specific heat in an anti-de Sitter 
space-time. The parametric relations for the occurence of Hawking-Page phase transitions and different
bounds on the parameters are determined. The exact extremality condition(\ref{extremality}) for the Ads Reissner-Nordstrom black hole
has been determined in terms of the mass, charge and cosmological constant. Our study specially emphasizes the
role of cosmological constant in deteminning the thermodynamic properties. 
Such cosmological constant may have its origin in a two brane generalized Randall-Sundrum model with bulk cosmological
constant appears through the parameter $k$ in the Physics on the brane. 
In the context of such a two brane generalized
Randall-Sundrum model, the existence of an upper bound in the value of the brane cosmological constant puts further 
constraints and bounds on the black hole parameters like mass and charge. We have derived  the 
exact correlation between the black hole mass and chrage with an inequality relation so that a horizon can be formed on the brane.    
Our work thus brings out various interesting details of the thermodynamics of black holes in an anti-de Sitter space-time,
embedded in a background 5-dimensional warped braneworld model. 
 
\acknowledgements{JM acknowledges Council for Scientific and
Industrial Research ( CSIR), Govt. of India for providing financial support.
RK thank the Department of Science and Technology (DST), Government
of India, for financial support through the project no.
SR/FTP/PS-68/2007. RK also thanks IACS where the major part of the
project has been done.}

\end{document}